\documentclass[10pt,twocolumn]{article}

\usepackage[
  left=1.55cm, right=1.55cm,
  top=2.0cm,   bottom=2.2cm,
  columnsep=0.65cm,
  headheight=14pt
]{geometry}

\usepackage{times}
\usepackage{mathptmx}
\usepackage[T1]{fontenc}
\usepackage[utf8]{inputenc}

\usepackage{amsmath,amssymb,amsfonts}
\usepackage{bm}

\usepackage{graphicx}
\usepackage{color}
\usepackage{tabularx}

\usepackage{microtype}
\usepackage{setspace}
\usepackage{parskip}
\usepackage{authblk}

\usepackage{xr-hyper} 
\usepackage{titlesec}
\titleformat{\section}[block]
  {\normalfont\bfseries\normalsize}{\thesection.}{0.5em}{}
\titlespacing{\section}{0pt}{8pt plus 2pt minus 1pt}{3pt plus 1pt}
\titleformat{\subsection}[runin]
  {\normalfont\bfseries\itshape\normalsize}{}{0pt}{}[.\quad]
\titlespacing{\subsection}{0pt}{6pt}{0pt}

\usepackage{caption}
\usepackage[numbers,sort&compress]{natbib}
\usepackage[breaklinks=true,colorlinks=true,allcolors=blue,
            pdftitle={Gate-tunable photoconductivity of graphene Josephson junctions}]{hyperref}

\usepackage{float}

\usepackage{fancyhdr}
\newcommand{\Ic}{I_{\mathrm{c}}}

\newcommand{\Rn}{R_{\mathrm{n}}}
\newcommand{\Te}{T_{\mathrm{e}}}

\newcommand{\Vph}{V_{\mathrm{ph}}}
\newcommand{\Pabs}{P_{\mathrm{abs}}}

\begin{document}

\title{
  \vspace{-1.8em}
  \Large\textbf{Gate-Tunable Photoresponse of Graphene Josephson Junctions
  at Terahertz Frequencies}
  \vspace{-0.8em}
}

\author[1,2]{Xiangyu Zhou}
\author[3]{Igor Gayduchenko}
\author[1,2]{Andrei Kudriashov}
\author[3]{Kirill Shein}
\author[4]{Anton Kuksov}
\author[1,2]{Leonid Elesin}
\author[1,2]{Mikhail Kravtsov}
\author[1,2]{Artur Shilov}
\author[1,2]{Olga Popova}
\author[1,2]{Subhajit Jana}
\author[5,6]{Takashi~Taniguchi}
\author[5,6]{Kenji~Watanabe}
\author[3]{Gregory Goltsman}
\author[1,2]{Kostya Novoselov}
\author[1,2]{Denis Bandurin*}

\affil[1]{Department of Materials Science and Engineering,
  National University of Singapore, 117575, Singapore}
\affil[2]{Institute for Functional Intelligent Materials,
  National University of Singapore, Singapore 117575, Singapore}
  \affil[3]{Moscow Pedagogical State University, Moscow 119991}
  \affil[4]{Programmable Functional Materials Lab,
Center for Neurophysics and Neuromorphic Technologies, Moscow, 121205, Russia}
\affil[5]{International Center for Materials Nanoarchitectonics,
  National Institute of Materials Science, Tsukuba 305-0044, Japan}
\affil[6]{Research Center for Functional Materials,
  National Institute of Materials Science, Tsukuba 305-0044, Japan}

\date{}
\maketitle
\thispagestyle{fancy}

\begin{abstract}
\noindent
Graphene Josephson junctions (JJ) provide a promising platform for ultra-broadband quantum sensing of light owing to graphene's frequency-independent absorption, vanishing electronic heat capacity, and weak electron--phonon coupling, which enable rapid suppression of the critical current through radiation-induced electron heating. Existing investigations have been confined to the microwave and infrared regimes, where competing detector technologies are already established; by contrast, the terahertz (THz) band - where sensitivity is most urgently lacking and no mature quantum sensor exists - has remained largerly unexplored. Here we demonstrate a strong photoresponse of graphene JJs at THz frequencies, establishing a first experimental step towards graphene-based THz quantum sensors. Under low-intensity illumination, we observe a pronounced suppression of the critical current that generates a strong photovoltage ($V_{\rm ph}$) under current bias. By tracking this $V_{\rm ph}$ and independently measuring the electron temperature as a function of absorbed power, we extract a responsivity of $88~\text{kV\,W}^{-1}$ and a noise-equivalent power of $45~\text{aW\,Hz}^{-1/2}$ at $1.7~\text{K}$. Furthermore, gate tunability of our JJ enables access to a regime where hysteretic current–voltage characteristics persist up to $0.9~\text{K}$, offering a potential route toward single-photon THz detection beyond millikelvin (mK) temperatures. These findings establish graphene JJ as a versatile platform for broadband cryogenic radiation sensing and point towards their use as quantum sensors at THz frequencies.

\end{abstract}


\vspace{4pt}
\noindent\rule{\columnwidth}{0.5pt}
\vspace{2pt}


\begin{figure*}[ht!]
\centering
\includegraphics[width=1\linewidth]{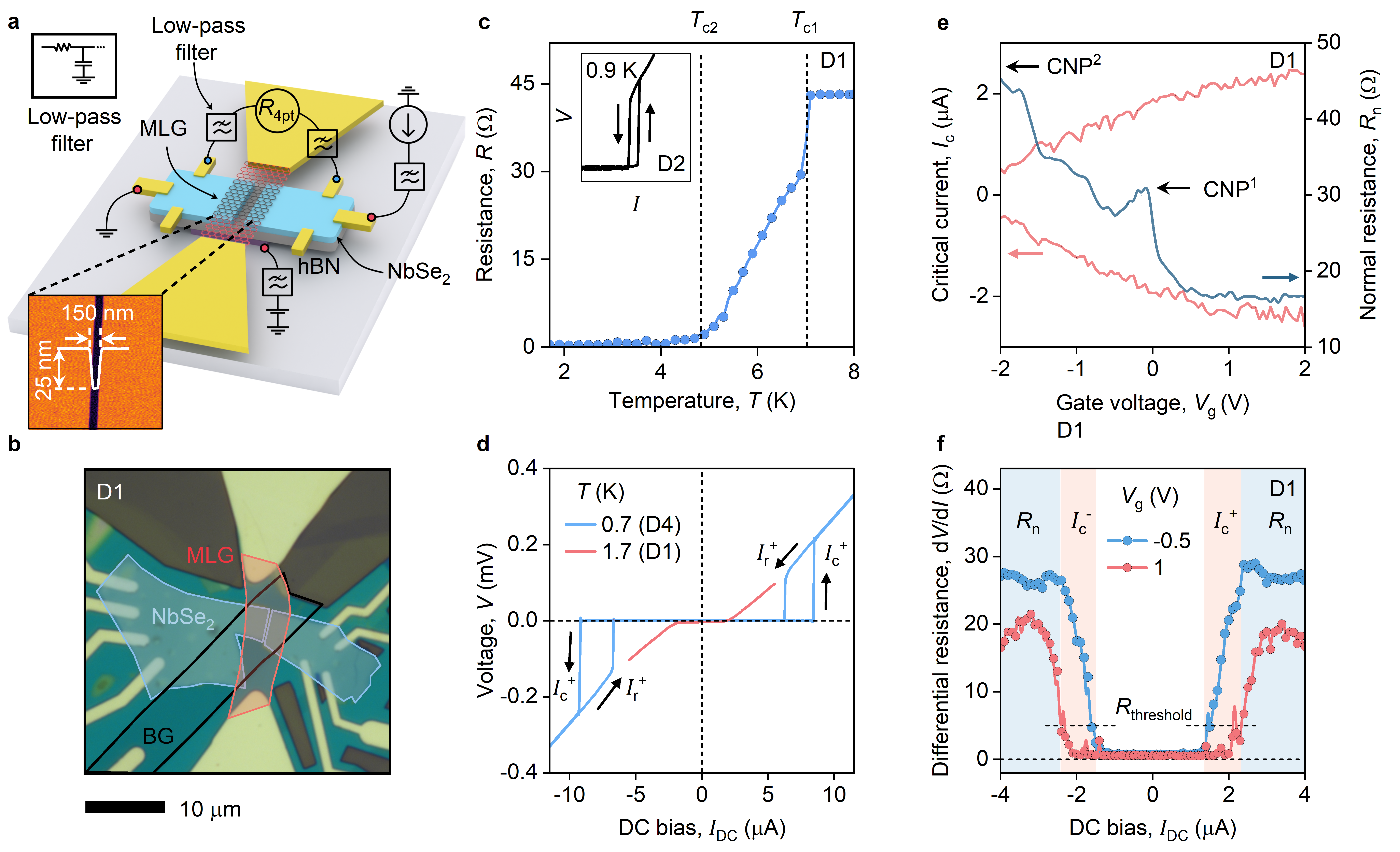}
\caption{%
  \textbf{Device architecture and superconducting transport.}
  \textbf{(a)}~Three-dimensional schematic of the van der Waals graphene JJ and the
  four-terminal measurement circuit. 
  The inset shows an AFM image of a representative
  cracked NbSe$_2$ electrode; the gap width is $\approx$150~nm and the flake
  thickness $\approx$25~nm.
  \textbf{(b)}~Optical micrograph of the main device, D1 in this study. The heterostructure consists, from top to bottom, of hBN, monolayer graphene (MLG), cracked NbSe$_2$, hBN,
  and graphite. 
  \textbf{(c)}~Temperature dependence of the four-terminal junction resistance of D1
  (at $I_{\rm DC}=0$). $T_\mathrm{c1}\approx 7$~K marks the bulk NbSe$_2$ transition; below $T_\mathrm{c2} \approx 4.8 $~K the resistance vanishes. The inset shows the hysteretic I--V behavior of the D2 at 0.9~K.
  \textbf{(d)}~Current--voltage ($I$--$V$) characteristics at 0.7~K (D4) and 1.7~K (D1).
  The hystersis between switching current $I_c$ and
  retrapping current $I_r$ at 0.7~K is signature of underdamped Josephson dynamics driven by self-Joule heating. At 1.7~K the hysteresis is absent. For clarity, the $I$–$V$ curve at 0.7~K shown here was measured on Device 4 (D4); all other data are from D1 unless otherwise specified.
  \textbf{(e)}~Critical current $\Ic$ (Red curve) and normal-state resistance $\Rn$ (Blue curve) as a function
  of gate voltage $V_g$ at $T = 1.7 $~K. CNP$^1$ (CNP$^2$) denotes the charge-neutrality point of
  peripheral MLG and MLG within the junction, respectively.
  \textbf{(f)} The differential resistance (d$V$/d$I$) as a function of DC bias, $I_{\rm DC}$. The d$V$/d$I$ curve shows robust superconducting pleatue with large crticial current and clear gate-tunable $I_{\rm c}$. The $I_{\rm c}$ can be extracted by setting a threshold resistance.
}
\label{fig:device}
\end{figure*}

\begin{figure*}[ht]
\centering
\includegraphics[width=\linewidth]{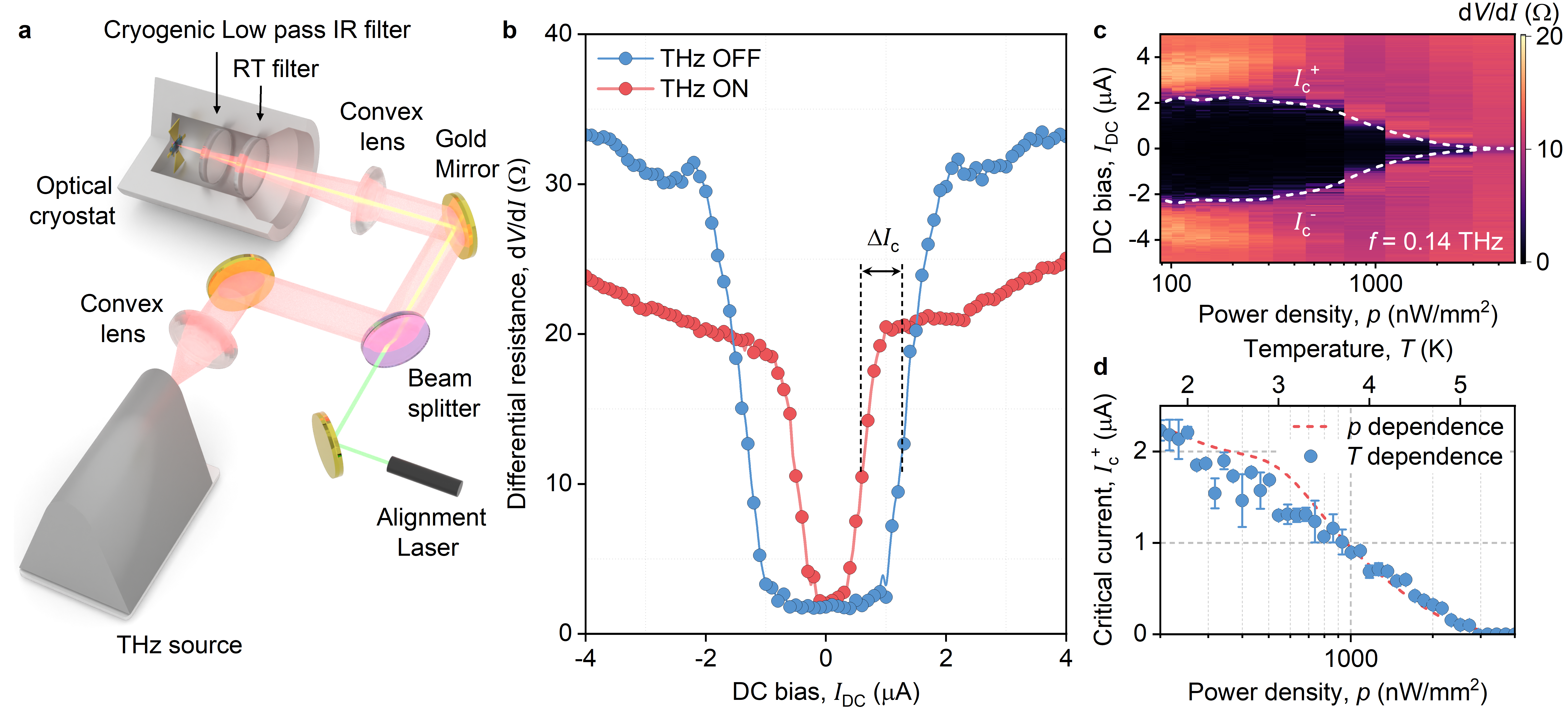}
\caption{%
  \textbf{THz photoresponse and power-dependent suppression of the critical current.}
  \textbf{(a)}~Optical setup. THz source is collimated by convex lens, co-aligned with a visible guide beam, and focused onto the sample by scanning a convex lens in $x$--$y$ to maximize the $V_{\rm ph}$ signal.
  IR-blocking filters at 300~K and 50~K suppress spurious blackbody loading.
  \textbf{(b)}~Differential resistance ${\rm d}V/{\rm d}I$ versus DC bias with
  (red) and without (blue) THz illumination at $V_g = 2$~V. THz irradiation produces a well-defined suppression $\Delta\Ic$ of the critical current.
  \textbf{(c)}~Colour map of ${\rm d}V/{\rm d}I$ versus $I_{\rm DC}$ and THz power
  density. The superconducting window bounded by the white dashed lines ($\Ic^{\pm}$) narrows symmetrically with increasing power and eventually disappers.
  \textbf{(d)}~Extracted $\Ic(P)$ versus THz power density at 1.7 K with the independently measured dark calibration curve $\Ic(T)$.
}
\label{fig:photoresponse}
\end{figure*}


The THz and millimeter (MM)-wave spectral ranges contain a substantial fraction of the Universe's radiative energy and are central to astronomy\cite{akiyama2019first, benford2004mission, graham2015experimental}, molecular spectroscopy\cite{jorgensen2012detection} and quantum technologies\cite{govia2015high}. Detecting this radiation is intrinsically challenging due to extremely small photon energies, demanding exceptional sensitivity, speed, and low noise\cite{siegel2002terahertz, li2025terahertz}. Cryogenic bolometric detectors can address some of these requirements by converting absorbed radiation into measurable electrical signals through temperature-dependent conductivity. Yet, modern detectors face a fundamental trade-off between sensitivity and response time, as both depend inversely on thermal conductivity~\cite{richards1994bolometers}. Overcoming this trade-off requires re-engineering the thermal degrees of freedom to drastically reduce effective heat capacity while maintaining resistive readouts. Hot-electron bolometers (HEBs) achieve this by confining heating to the electronic subsystem rather than the entire device structure\cite{li2025terahertz}. For example, superconducting HEBs feature nanosecond-scale response through radiation-induced heating of electrons near the critical temperature, yet their responsivity generally remains far below semiconducting electron bolometers~\cite{sizov2018terahertz}. Two-dimensional electron systems offer a promising alternative route~\cite{shein2024fundamental}. Graphene, in particular, combines ultralow electronic specific heat\cite{aamir2021ultrasensitive} with rapid electron–electron thermalization \cite{brida2013ultrafast} and weak electron–phonon coupling, allowing the electronic temperature to rise sharply above the lattice temperature under minute absorbed power\cite{tielrooij2015generation, borzenets2013phonon, betz2013supercollision, efetov2018fast, fong2012ultrasensitive} and cool down on a picosecond scale after the excitation is removed~\cite{graham2013photocurrent, block2021observation,massicotte2021hot}. However, graphene's single-particle conductivity depends only weakly on electronic temperature, severely limiting its use as a standalone bolometric sensor and motivating the development of more sophisticated device architectures to convert electronic heating into a measurable electrical signal\cite{vora2012bolometric, yan2012dual, wei2008ultrasensitive, efetov2018fast, han2013highly, lee2020graphene, walsh2017graphene, sarkar2025kerr, kokkoniemi2020bolometer, walsh2021josephson, kravtsov2025viscous, el2016epitaxial, riccardi2020ultrasensitive, han2013highly, lara2019towards, bandurin2018dual, yan2012dual}. 

One of the most promising architectures for this enquiry involves integrating graphene as a weak link in Josephson junction (JJ) where radiation-induced change of the critical current provides measurable signal. Although graphene JJs featured ultra-sensitive detection of microwaves and infrared (IR) radiation with single-photon sensitivity\cite{walsh2017,kokkoniemi2020bolometer, lee2020graphene,sarkar2025kerr, walsh2021,huang2026thermal}, their response at THz frequencies remained largely unexplored, with only one exception based on JJ made of bilayer graphene on silicon carbide~\cite{CarbonJJ}. The latter approach was, however, limited by diffusive ungated device geometry, the larger electronic heat capacity of the BLG link, the need for SQUID-based readout at mK temperatures, and relied on spectrally filtered (effectively single-frequency) blackbody radiation. In this work, we address these limitations and demonstrate giant tunable THz photoresponse of graphene JJ at $1.7~$K and with functional response up to liquid helium temperatures spanning across MM-wave to far-IR spectral ranges. Finally, by electrostatically tuning our devices into the underdamped regime reveals hysteretic current-voltage characteristics that persist up to $900~\text{mK}$, which - combined with a picosecond energy-relaxation time - opens a pathway towards THz quantum sensing.


\section*{Device Architecture and Superconducting Transport}

Figure~\ref{fig:device}a shows a schematic of the van der Waals (vdW) graphene JJ. Monolayer graphene (MLG) serves as the weak link between two 
superconducting contacts formed by a naturally cracked NbSe$_2$ flake, whose crack 
defines the junction gap and yields an atomically clean interface, as confirmed by 
the AFM image in the inset of Fig.~\ref{fig:device}a (exfolliation, vdW assembly and AFM investigation were conducted in an inert environment of argon-filled glovebox). Both the MLG and NbSe$_2$  are encapsulated between hexagonal boron nitride (hBN) layers, which preserve the 
flatness of the heterostructure and protect it from environmental degradation. A 
graphite flake beneath the lower hBN serves as a local electrostatic gate, enabling 
continuous tuning of the carrier density in the graphene weak link. Electrical 
contacts are defined by standard electron-beam lithography followed by thermal 
evaporation of Ti/Au. To minimise electrical noise, low-pass RC filters are 
connected to all leads, and the device ground is thermally anchored to the cold 
finger of the optical cryostat. Figure~\ref{fig:device}b shows an optical micrograph 
of the representative device of the study, Device 1(D1).
The photoresponse of the devices was measured in Quantum Design Optical cryostat (1.7--350~K) while additional transport experiments were conducted in Bluefors dilution refrigerator (10~mK--1~K). 

Prior to photoresponse measurements, we first established the transport characteristics of the devices. Above 7 K, the junction’s normal resistance $R_\mathrm{0}$ is $43~\Omega$ and is nearly $T-$independent (Fig.~\ref{fig:device}c). As the temperature is reduced, $R_\mathrm{0}$ drops sharply at $T_\mathrm{c1}=7~$K, marking the onset of superconductivity in NbSe$_2$, and reaches zero at $T_\mathrm{c2}\approx 4.8$~K, where Andreev transport gives rise to a fully developed Josephson state across the graphene. Representative $I$--$V$ characteristics at 1.7~K and 700~mK for two representative devices are shown in Fig.~\ref{fig:device}d. Additional data, shown in the inset of Fig.~\ref{fig:device}c, demonstrates that switching hysteretic behavior in such junctions can persist up to 0.9~K. Below we focus only on one device labeled as D1;  data from other devices are shown in Supplementary Information. At $T=1.7$~K--the base temperature of our optical cryostat - the junction still supports a robust supercurrent, as evidenced by its zero-resistance state (See Fig.~\ref{fig:device}f for the differential resistance ${\rm d}V/{\rm d}I$ data). The Josephson effect in our devices is strongly gate-tunable: $I_\mathrm{c}$ decreases from $2\mu$A to $\sim 500$~nA as the carrier density approaches the charge neutrality point (Fig.~\ref{fig:device}e). We further observe that NbSe$_2$ induces doping in the graphene channel, shifting the neutrality point to $V_\mathrm{g}<-2$~V (labelled CNP$^2$). In addition, a pronounced resistance peak near $V_\mathrm{g}=0$~V (Fig.~\ref{fig:device}e) is attributed to current paths outside the proximitized NbSe$_2$ region. In Fig.~\ref{fig:device}f, two representative ${\rm d}V/{\rm d}I$ curves at different gate voltages $V_{\rm g}$ are presented. Both show robust superconductivity, while the difference in $I_{\rm c}$ arises from the variation in carrier density. The critical current $I_{\rm c}$ is determined using a resistance-threshold ($R_{\rm threshold}$) method.


\section*{Critical current suppression in THz-exposed graphene JJ}

Figure~\ref{fig:photoresponse}a shows the optical setup that we used to couple THz radiation to the graphene JJs. Continuous-wave radiation is generated either by an IMPATT diode source (Terasense, 0.14~THz) or by a quantum cascade laser (Lytid) operating at 2.5 and 3.5~THz. The THz beam is guided and focused onto the sample using a set of convex lenses and a gold mirror. A beam splitter directs a co-aligned visible laser beam, which is used to facilitate alignment of the optical path. The radiation enters an optical cryostat through a series of low-pass filters mounted at room temperature and at 50~K, which suppress unwanted visible and mid-infrared background while transmitting THz frequencies. Inside the cryostat, the beam is focused onto the device, ensuring efficient coupling of the incident radiation to the graphene Josephson junction.

\begin{figure*}[ht]
\centering
\includegraphics[width=\linewidth]{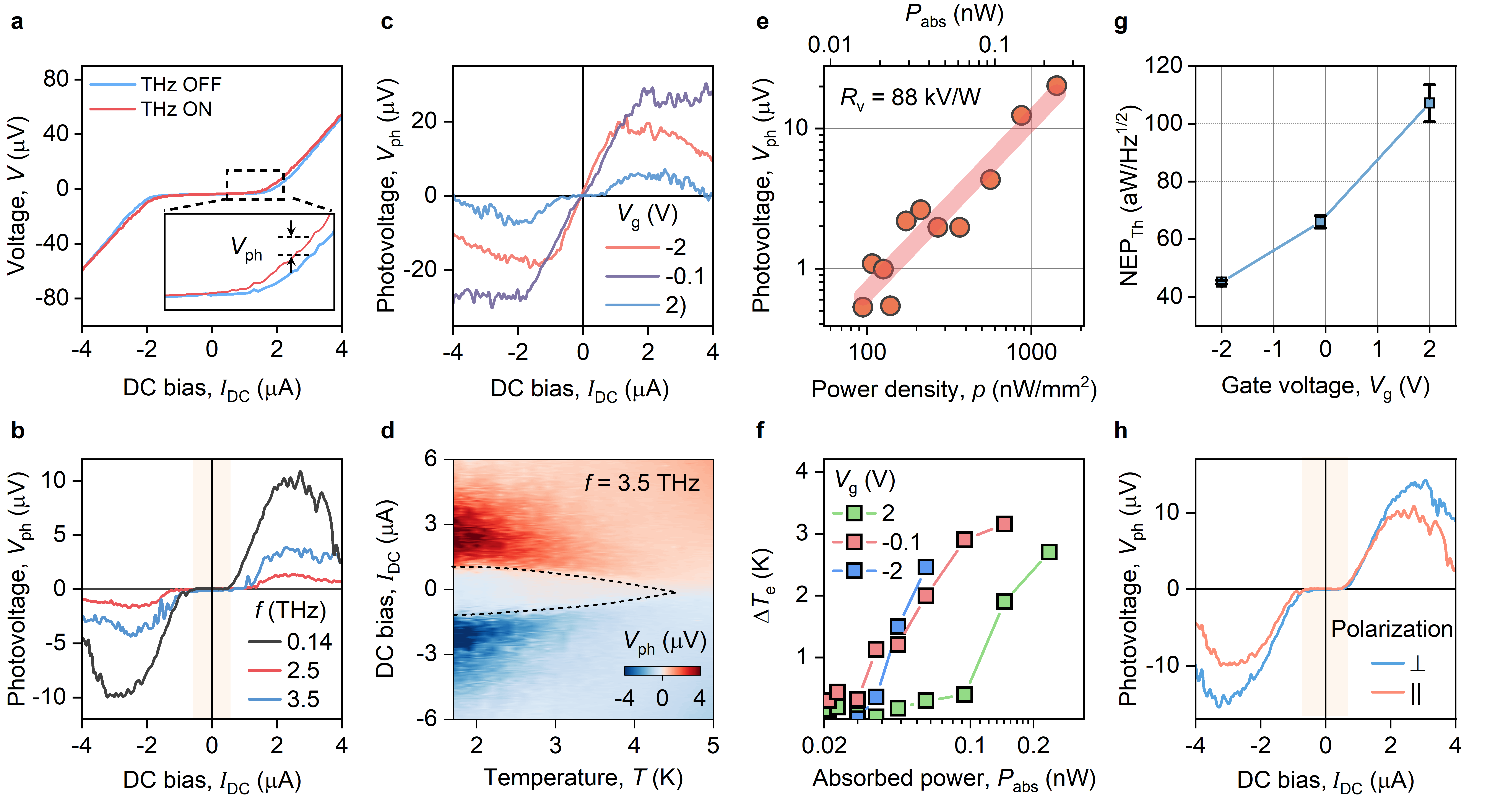}
\caption{
  \textbf{Photovoltage, responsivity, and heat-transport analysis.}
  \textbf{(a)} I-V characteristics of GJJ with THz (3.5~THz) ON and OFF. The suppression of $\Ic$ is demonstrated. The $V_{\rm ph} $ is shown in the inset. 
  \textbf{(b)}~$\Vph$ versus $I_{\rm DC}$
  measured at 0.14~THz, 2.5~THz, and 3.5~THz. All three frequencies produce qualitatively identical bias-current profiles; amplitude differences
  reflect the varying source output powers.
  \textbf{(c)}~$\Vph$ at 0.14~THz versus $I_{\rm DC}$ at three gate voltages for fixed incident THz power. The response is maximized near CNP$^1$, where the vanishing carrier density minimizes the electronic heat capacity of the peripheral MLG.
  \textbf{(d)}~$V_{\rm ph}$ at 3.5~THz versus $I_{\rm DC}$ and $T$. The amplitude decreases with increasing temperature, consistent with a Josephson origin. The region without $V_{\rm ph}$, bounded by dashed lines, corresponds to intact superconductivity under THz irradiation.
  \textbf{(e)}~$\Vph$ versus power density and corresponding calibrated absorbed power $\Pabs$ (see text). The linear fit gives a voltage responsivity $R_\mathrm{v} = 87.6$~kV~W$^{-1}$.
  \textbf{(f)}~Electron-temperature rise $\Delta\Te = \Te - T_{\rm }$ extracted from the $\Ic$ suppression, plotted versus $\Pabs$ at three gate voltages.
  The response is weakest in the heavily electron-doped regime ($V_g = +2$~V) where the heat capacity is largest.
  \textbf{(g)} The calculated theoretical noise-equivalent power (NEP) due to thermal fluctuations.
  \textbf{(h)}~Polarization dependence of $\Vph$: THz polarization parallel and perpendicular to the bow-tie antenna yield nearly equal amplitudes, indicating that direct free-carrier absorption in the graphene antenna wings dominates over
  resonant antenna coupling.
}
\label{fig:photovoltage}
\end{figure*}

Figure~\ref{fig:photoresponse}b shows ${\rm d}V/{\rm d}I$ as a function of current bias $I_\mathrm{DC}$ measured in the dark and upon exposure to 0.14~THz radiation. In the absence of irradiation, the junction exhibits a pronounced nearly zero-resistance region corresponding to the superconducting state. Under THz illumination, this region narrows significantly, reflecting a suppression of the critical current $I_\mathrm{c}$. This reduction is quantified by the shift $\Delta I_\mathrm{c}$ between the switching currents in the dark and illuminated states. This behavior is consistently observed for all accessible gate voltages (Supplementary Information).

To confirm that the observed effect originates from the incident THz radiation, we first investigate the dependence of $I_\mathrm{c}$ on the power density $p$ incident onto our device for the case of 0.14 THz excitation. To this end, in Fig.~\ref{fig:photoresponse}c we plot $dV/dI$ vs $p$: the superconducting window, bounded by the white dashed lines ($I_c^{\pm}$), narrows symmetrically with increasing $p$ and eventually collapses at high power levels. This behavior is consistent with thermal suppression of superconductivity: increasing THz power elevates the electronic temperature, which reduces the critical current and ultimately drives the junction into the normal state~\cite{lee2020graphene}. As increasing the cryostat temperature also leads to a reduction in $I_{\rm c}$, the electron temperature can be extracted by comparing the $I_{\rm c}$ under radiation with the dark $I_{\rm c}$ measured at different bath temperatures. The $I_{\rm c}$ under radiation at base temperature and the dark $I_{\rm c}$ at elevated temperatures are shown in Fig.~\ref{fig:photoresponse}d.


\section*{Photovoltage, Responsivity, and Noise Equivalent Power}

Having established that under THz illumination the critical current is suppressed, we turn back to the $I$--$V$ characteristics measured in the dark and under THz irradiation. As shown in Fig.~\ref{fig:photovoltage}a, the two curves coincide at zero bias. Upon increasing the current, the curves begin to deviate in the vicinity of the superconducting-to-normal transition. In this regime, THz-induced suppression of the critical current shifts the transition range, leading to a finite voltage difference between the illuminated and dark traces, which we define as the photovoltage ($V_{\rm ON } - V_{\rm OFF}$). At higher bias, once the junction is fully in the normal state, the two curves merge again, corresponds to the decrease of $V_{\rm ph}$. To measure it with high sensitivity, we modulate the THz radiation and employ standard lock-in detection. The resulting $V_\mathrm{ph}$ as a function of bias current is shown in Fig.~\ref{fig:photovoltage}c. The $V_{\rm ph}$ vanishes at zero bias, increases upon approaching the critical current, and decreases again in the normal state. Because the critical current depends on gate voltage $V_\mathrm{g}$, the $V_{\rm ph}$ is likewise gate-tunable, as demonstrated in Fig.~\ref{fig:photovoltage}c, where both the magnitude and shape of $V_\mathrm{ph}(I_\mathrm{DC})$ vary with $V_\mathrm{g}$.

We have also studied the response of our devices to THz radiation at different frequencies and find that the overall shape of the $V_{\rm ph}$ remains the same for 0.14, 2.5 and 3.5 THz. As shown in Fig.~\ref{fig:photovoltage}b, $V_\mathrm{ph}(I_\mathrm{DC})$ exhibits the same characteristic behavior for all excitation frequencies: it vanishes at zero bias, grows in magnitude near the superconducting transition, and decreases again in the normal state. The observed differences in amplitude arise from variations in the incident power density at different frequencies. These results demonstrate that the device operates robustly across a broad spectral range, spanning from MM-wave to far-IR frequencies.

Figure~3d shows $V_{\mathrm{ph}}$ of our device as a function of $I_{\mathrm{DC}}$ and temperature $T$ under illumination with $f = 3.5~\mathrm{THz}$ radiation. The map reveals an expected antisymmetric response with respect to current bias: positive photovoltage (red) appears for positive $I_{\mathrm{DC}}$, while negative photovoltage (blue) develops for negative $I_{\mathrm{DC}}$. The magnitude of the signal is strongest at low temperatures and close to the critical current, reaching several $\mu\mathrm{V}$, and gradually diminishes with increasing temperature, becoming nearly suppressed above $5~\mathrm{K}$. This behavior indicates that the photoresponse is closely linked to the superconducting state of the junction and its temperature-dependent critical current.

To quantify the detector performance, we extract the maximum $V_{\rm ph}$ from the $V_\mathrm{ph}(I_\mathrm{DC})$ curves and plot it as a function of incident power density at 0.14~THz, as shown in Fig.~\ref{fig:photovoltage}e. The device exhibits a wide dynamic range; however, focusing on the linear regime allows us to define a well-defined responsivity. To convert incident power to absorbed power, we employ a calibration procedure based on electron heating. Specifically, we exploit the sensitivity of the graphene resistance near the charge neutrality point to the electronic temperature $T_\mathrm{e}$. By applying a DC current to induce Joule heating, we monitor the corresponding change in resistance and compare it to the change induced by THz illumination. This establishes a one-to-one correspondence between incident and absorbed power (see Supplementary Information for details). Using this calibration, we plot $V_\mathrm{ph}$ as a function of absorbed power $P_\mathrm{abs}$ in Fig, \ref{fig:photovoltage}e and extract the responsivity from the slope in the linear regime, obtaining $R_V \approx 88~\text{kV\,W}^{-1}$. This value by order of magnitude larger than that of previously reported superconducting hot-electron bolometers.

It is also instructive to estimate the noise-equivalent power, $\mathrm{NEP}_\mathrm{th}$, which is governed by the thermal fluctuations and expressed throught the thermal conductance $G_\mathrm{th}$ as $\mathrm{NEP}_\mathrm{th} = \sqrt{4 k_\mathrm{B} T_\mathrm{e}^2 G_\mathrm{th}}$. To determine $G_\mathrm{th}$, we relate the increase in electronic temperature $\Delta T_\mathrm{e}$ to the absorbed power $P_\mathrm{abs}$. This can be achieved by returning to Fig.~\ref{fig:photoresponse}d. By monitoring how the critical current $I_\mathrm{c}$ evolves with incident power and comparing it to its independent temperature dependence, we establish a correspondence between $P_\mathrm{abs}$ and $\Delta T_\mathrm{e}$, assuming that the response is governed by electron heating. From this relation, we extract $\Delta T_\mathrm{e}(P_\mathrm{abs})$ and determine the thermal conductance at the operating point (finite $\Delta T_\mathrm{e}$) from the slope, yielding the dynamic $G_\mathrm{th} = dP_\mathrm{abs}/dT_\mathrm{e}$. Using this value, we estimate $\mathrm{NEP}_\mathrm{th}$, as shown in Fig.~\ref{fig:photovoltage}g. The extracted $\mathrm{NEP}_\mathrm{th}=45-100$ aW$/\sqrt{Hz}$ is within the range of the previous report~\cite{CarbonJJ}, while achieved here at temperatures an order of magnitude higher and without involving SQUID measurements. Furthermore, $\mathrm{NEP}_\mathrm{th}$ can be tuned via the gate voltage, reaching its optimum at $V_\mathrm{g} = -2~\mathrm{V}$, where the critical current is minimized.

\section*{Discussion}


The excellent responsivity, low theoretical NEP, gate tunability, broadband operation, and potentially ultra-fast response (see Ref.~\cite{titova2026fast}) distinguish our graphene Josephson junctions from existing superconducting THz detectors. Nevertheless, it is important to discuss the limitations of the present device architecture and identify possible routes for improvement. First, we note that our devices are intentionally equipped with a broadband bow-tie antenna designed to couple THz radiation to graphene regions adjacent to the junction. However, polarization-dependent measurements of $V_\mathrm{ph}(I_\mathrm{DC})$ (Fig.~\ref{fig:photovoltage}h) reveal nearly identical responses for two orthogonal polarizations, indicating that the antenna is effectively inactive. We attribute this behavior to strong impedance mismatch between the antenna and the graphene junction, resulting in inefficient coupling. The absence of controlled antenna coupling complicates the estimation of the extrinsic responsivity, as the effective absorption area is not well defined. If one assumes that the incident THz radiation is absorbed uniformly over the graphene regions outside the junction, one obtains an estimated responsivity on the order of $\sim 100~\text{kV\,W}^{-1}$, consistent with intrinsic calibration.

\begin{figure}[htbp]
    \centering
    \includegraphics[width=\columnwidth]{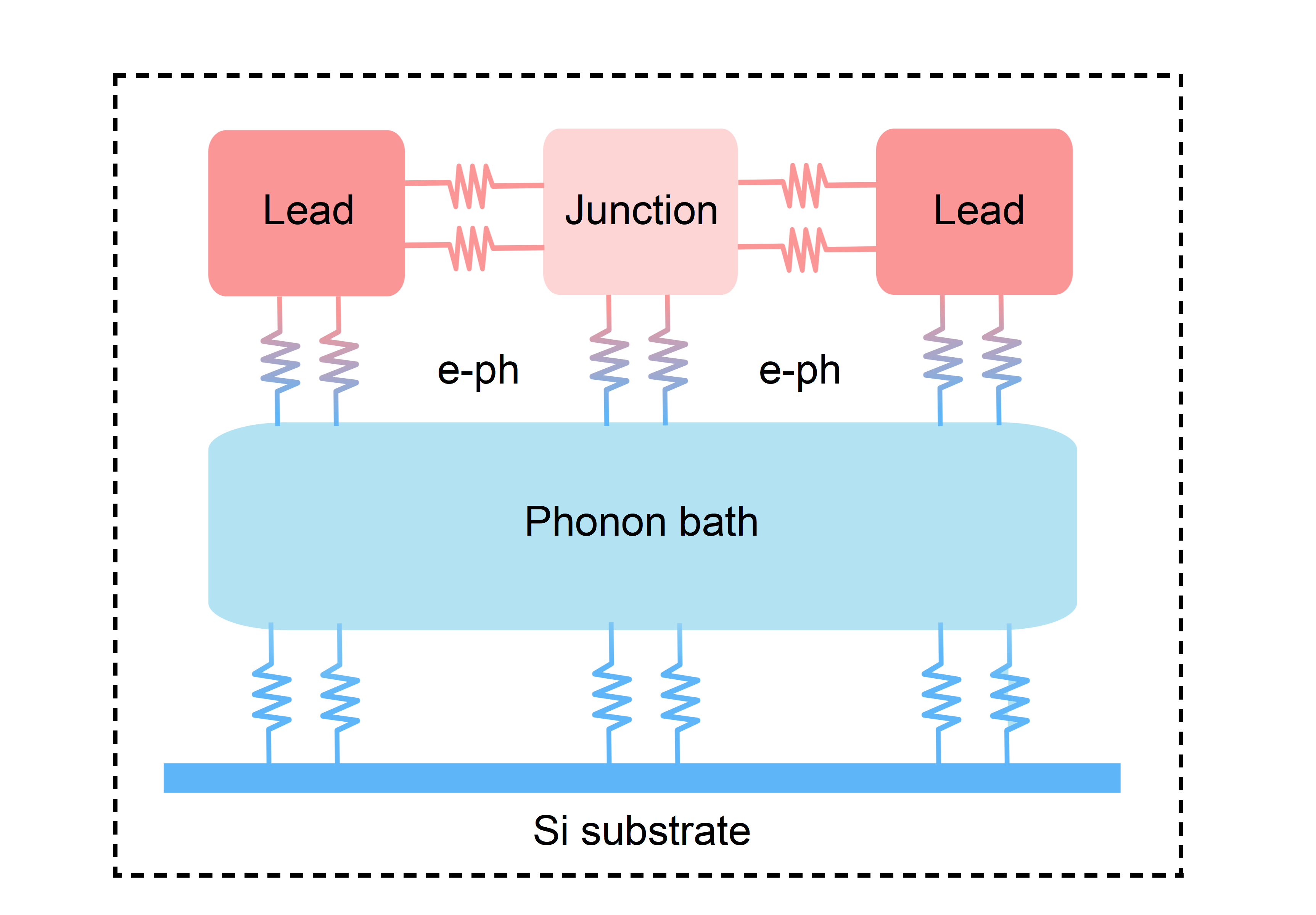}
    \caption{The schematic of heat transport in our graphene JJ. Heat generated in graphene leads diffuses to the junction to and to phonons through electron-phonon coupling.}
    \label{fig:discussion}
\end{figure}

Importantly, this scenario implies that the dominant heat injection occurs in graphene regions spatially separated from the junction, with energy subsequently reaching the junction via hot-electron diffusion. This picture is illustrated in the schematic in Fig.~\ref{fig:discussion}, where the graphene leads act as extended absorbers thermally coupled to the junction via electronic diffusion and to the phonon bath via electron--phonon interactions. Such a mechanism is consistent with recent demonstrations of graphene-based single-photon detectors, where thermal hotspots generated several microns away from the junction were shown to trigger switching events~\cite{huang2026thermal}. In those systems, the long thermal diffusion length of Dirac electrons enables efficient energy transport across the device. 

These observations suggest clear pathways for improvement and next steps. First, impedance-matched antenna designs or cavity integration could enable efficient and localized coupling directly to the junction. Second, engineering thermal isolation could enhance the temperature rise and responsivity. Third, our graphene Josephson junction devices exhibit hysteretic behavior at temperatures as high as 900~mK (see Supplementary Information), which motivates extending THz optoelectronic measurements to this regime and probing switching upon decreasing incident THz power to explore the possibility of single-photon detection. Finally, controlled positioning of the absorption region relative to the junction may allow optimization of the trade-off between responsivity and speed.

\section*{Conclusion}

In summary, we have demonstrated that graphene Josephson junctions exhibit a strong, gate-tunable photoresponse at terahertz frequencies, establishing them as a promising platform for broadband cryogenic THz sensing. Under low-intensity illumination, THz radiation suppresses the critical current and generates a measurable photovoltage under current bias. By calibrating the absorbed power through independent electron-heating measurements, we extract a responsivity of about $88~\mathrm{kV\,W^{-1}}$ and a theoretical noise-equivalent power in the $45$--$100~\mathrm{aW\,Hz^{-1/2}}$ range at $1.7~\mathrm{K}$. The response persists from millimeter-wave to far-infrared frequencies and can be tuned electrostatically through the gate, highlighting the versatility of this architecture. Together with the observation of hysteretic Josephson dynamics up to $900~\mathrm{mK}$ and the intrinsically fast energy-relaxation times of graphene, these results identify graphene Josephson junctions as a compelling route toward ultrasensitive and potentially single-photon THz detectors. Beyond the present device design, our findings also define clear directions for improvement through optimized coupling and better control of heat flow, providing a practical roadmap toward graphene-based THz quantum sensors.




\bibliography{bibliography}

@article{kokkoniemi2020bolometer,
  title={Bolometer operating at the threshold for circuit quantum electrodynamics},
  author={Kokkoniemi, Roope and Girard, J-P and Hazra, Dibyendu and Laitinen, Antti and Govenius, Joonas and Lake, RE and Sallinen, Iiro and Vesterinen, Visa and Partanen, Matti and Tan, JY and others},
  journal={Nature},
  volume={586},
  number={7827},
  pages={47--51},
  year={2020},
  publisher={Nature Publishing Group UK London}
}

@misc{titova2026fast,
      title={Graphene Zero-Bias Sub-Terahertz Turnkey Detector with Above 43 GHz Bandwidth}, 
      author={E. I. Titova and A. Titchenko and M. Titova and K. Shein and A. Kuksov and A. Sobolev and M. Kashchenko and M. Kravtsov and L. Elesin and K. S. Novoselov and G. Goltsman and D. A. Svintsov and I. Gayduchenko and D. A. Bandurin},
      year={2026},
      eprint={2603.03554},
      archivePrefix={arXiv},
      primaryClass={cond-mat.mes-hall},
      url={https://arxiv.org/abs/2603.03554}, 
}

@article{CarbonJJ,
title = {A terahertz detector based on superconductor-graphene-superconductor Josephson junction},
journal = {Carbon},
volume = {202},
pages = {112-117},
year = {2023},
issn = {0008-6223},
}

@article{sarkar2025kerr,
  title={Kerr non-linearity enhances the response of a graphene Josephson bolometer},
  author={Sarkar, Joydip and Maji, Krishnendu and Sunamudi, Abhishek and Agarwal, Heena and Samanta, Priyanka and Bhattacharjee, Anirban and Rajkhowa, Rishiraj and Patankar, Meghan P and Watanabe, Kenji and Taniguchi, Takashi and others},
  journal={Nature Communications},
  volume={16},
  number={1},
  pages={7043},
  year={2025},
  publisher={Nature Publishing Group UK London}
}

@article{huang2026thermal,
  title={Thermal detection of single photons using Dirac fermions},
  author={Huang, Bevin and Arnault, Ethan G and Jung, Woochan and Fried, Caleb and Russell, B Jordan and Watanabe, Kenji and Taniguchi, Takashi and Henriksen, Erik A and Englund, Dirk and Lee, Gil-Ho and others},
  journal={Nature Communications},
  year={2026},
  publisher={Nature Publishing Group UK London}
}

@article{walsh2017graphene,
  title={Graphene-based Josephson-junction single-photon detector},
  author={Walsh, Evan D and Efetov, Dmitri K and Lee, Gil-Ho and Heuck, Mikkel and Crossno, Jesse and Ohki, Thomas A and Kim, Philip and Englund, Dirk and Fong, Kin Chung},
  journal={Physical Review Applied},
  volume={8},
  number={2},
  pages={024022},
  year={2017},
  publisher={APS}
}

@article{lara2019towards,
  title={Towards quantum-limited coherent detection of terahertz waves in charge-neutral graphene},
  author={Lara-Avila, S and Danilov, A and Golubev, D and He, Hans and Kim, KH and Yakimova, Rositsa and Lombardi, F and Bauch, T and Cherednichenko, S and Kubatkin, S},
  journal={Nature Astronomy},
  volume={3},
  number={11},
  pages={983--988},
  year={2019},
  publisher={Nature Publishing Group UK London}
}

@article{shein2024fundamental,
  title={Fundamental limits of few-layer NbSe2 microbolometers at terahertz frequencies},
  author={Shein, Kirill and Zharkova, Ekaterina and Kashchenko, Mikhail and Kolbatova, Anna and Lyubchak, Anastasia and Elesin, Leonid and Nguyen, Ekaterina and Semenov, Alexander and Charaev, Ilya and Schilling, Andreas and others},
  journal={Nano letters},
  volume={24},
  number={7},
  pages={2282--2288},
  year={2024},
  publisher={ACS Publications}
}

@article{li2025terahertz,
  title={Terahertz science and technology in astronomy, telecommunications, and biophysics},
  author={Li, Jing and Deng, Xianjin and Li, Yangmei and Hu, Jie and Miao, Wei and Lin, Changxing and Jiang, Jun and Shi, Shengcai},
  journal={Research},
  volume={8},
  pages={0586},
  year={2025},
  publisher={AAAS}
}

@article{akiyama2019first,
  title={First M87 event horizon telescope results. IV. Imaging the central supermassive black hole},
  author={Akiyama, Kazunori and Alberdi, Antxon and Alef, Walter and Asada, Keiichi and Azulay, Rebecca and Baczko, Anne-Kathrin and Ball, David and Balokovi{\'c}, Mislav and Barrett, John and Bintley, Dan and others},
  journal={The Astrophysical Journal Letters},
  volume={875},
  number={1},
  pages={L4},
  year={2019},
  publisher={IoP Publishing}
}

@article{jorgensen2012detection,
  title={Detection of the simplest sugar, glycolaldehyde, in a solar-type protostar with ALMA},
  author={J{\o}rgensen, Jes K and Favre, C{\'e}cile and Bisschop, Suzanne E and Bourke, Tyler L and Van Dishoeck, Ewine F and Schmalzl, Markus},
  journal={The Astrophysical Journal Letters},
  volume={757},
  number={1},
  pages={L4},
  year={2012},
  publisher={IOP Publishing}
}

@article{tielrooij2015generation,
  title={Generation of photovoltage in graphene on a femtosecond timescale through efficient carrier heating},
  author={Tielrooij, Klaas-Jan and Piatkowski, Lukasz and Massicotte, Mathieu and Woessner, Achim and Ma, Qiong and Lee, Yongjin and Myhro, Kevin Scott and Lau, Chun Ning and Jarillo-Herrero, Pablo and van Hulst, Niek F and others},
  journal={Nature nanotechnology},
  volume={10},
  number={5},
  pages={437--443},
  year={2015},
  publisher={Nature Publishing Group UK London}
}

@article{richards1994bolometers,
  title={Bolometers for infrared and millimeter waves},
  author={Richards, Paul L},
  journal={Journal of Applied Physics},
  volume={76},
  number={1},
  pages={1--24},
  year={1994},
  publisher={American Institute of Physics}
}

@article{siegel2002terahertz,
  title={Terahertz technology},
  author={Siegel, Peter H},
  journal={IEEE Transactions on microwave theory and techniques},
  volume={50},
  number={3},
  pages={910--928},
  year={2002},
  publisher={IEEE}
}

@article{aamir2021ultrasensitive,
  title={Ultrasensitive calorimetric measurements of the electronic heat capacity of graphene},
  author={Aamir, Mohammed Ali and Moore, John N and Lu, Xiaobo and Seifert, Paul and Englund, Dirk and Fong, Kin Chung and Efetov, Dmitri K},
  journal={Nano Letters},
  volume={21},
  number={12},
  pages={5330--5337},
  year={2021},
  publisher={ACS Publications}
}

@article{sizov2018terahertz,
  title={Terahertz radiation detectors: the state-of-the-art},
  author={Sizov, F},
  journal={Semiconductor science and technology},
  volume={33},
  number={12},
  pages={123001},
  year={2018},
  publisher={IOP Publishing}
}

@article{vora2012bolometric,
  title={Bolometric response in graphene based superconducting tunnel junctions},
  author={Vora, Heli and Kumaravadivel, Piranavan and Nielsen, Bent and Du, Xu},
  journal={Applied Physics Letters},
  volume={100},
  number={15},
  year={2012},
  publisher={AIP Publishing}
}

@article{yan2012dual,
  title={Dual-gated bilayer graphene hot-electron bolometer},
  author={Yan, Jun and Kim, Mann Ho and Elle, Jennifer A and Sushkov, Andrei B and Jenkins, Greg S and Milchberg, Howard M and Fuhrer, Michael S and Drew, HD},
  journal={Nature nanotechnology},
  volume={7},
  number={7},
  pages={472--478},
  year={2012},
  publisher={Nature Publishing Group UK London}
}

@article{han2013highly,
  title={Highly sensitive hot electron bolometer based on disordered graphene},
  author={Han, Qi and Gao, Teng and Zhang, Rui and Chen, Yi and Chen, Jianhui and Liu, Gerui and Zhang, Yanfeng and Liu, Zhongfan and Wu, Xiaosong and Yu, Dapeng},
  journal={Scientific reports},
  volume={3},
  number={1},
  pages={3533},
  year={2013},
  publisher={Nature Publishing Group UK London}
}

@article{borzenets2013phonon,
  title={Phonon bottleneck in graphene-based Josephson junctions at millikelvin temperatures},
  author={Borzenets, IV and Coskun, UC and Mebrahtu, HT and Bomze, Yu V and Smirnov, AI and Finkelstein, Gleb},
  journal={Physical review letters},
  volume={111},
  number={2},
  pages={027001},
  year={2013},
  publisher={APS}
}

@article{betz2013supercollision,
  title={Supercollision cooling in undoped graphene},
  author={Betz, AC and Jhang, Sung Ho and Pallecchi, Emiliano and Ferreira, Robson and F{\`e}ve, Gwendal and Berroir, Jean-Marc and Pla{\c{c}}ais, Bernard},
  journal={Nature Physics},
  volume={9},
  number={2},
  pages={109--112},
  year={2013},
  publisher={Nature Publishing Group UK London}
}

@article{wei2008ultrasensitive,
  title={Ultrasensitive hot-electron nanobolometers for terahertz astrophysics},
  author={Wei, Jian and Olaya, David and Karasik, Boris S and Pereverzev, Sergey V and Sergeev, Andrei V and Gershenson, Michael E},
  journal={Nature nanotechnology},
  volume={3},
  number={8},
  pages={496--500},
  year={2008},
  publisher={Nature Publishing Group UK London}
}

@article{bandurin2018dual,
  title={Dual origin of room temperature sub-terahertz photoresponse in graphene field effect transistors},
  author={Bandurin, DA and Gayduchenko, I and Cao, Y and Moskotin, M and Principi, Alessandro and Grigorieva, IV and Goltsman, G and Fedorov, G and Svintsov, D},
  journal={Applied Physics Letters},
  volume={112},
  number={14},
  year={2018},
  publisher={AIP Publishing}
}

@article{kravtsov2025viscous,
  title={Viscous terahertz photoconductivity of hydrodynamic electrons in graphene},
  author={Kravtsov, M and Shilov, AL and Yang, Y and Pryadilin, T and Kashchenko, MA and Popova, O and Titova, M and Voropaev, D and Wang, Y and Shein, K and others},
  journal={Nature Nanotechnology},
  volume={20},
  number={1},
  pages={51--56},
  year={2025},
  publisher={Nature Publishing Group UK London}
}

@article{brida2013ultrafast,
  title={Ultrafast collinear scattering and carrier multiplication in graphene},
  author={Brida, Daniele and Tomadin, Andrea and Manzoni, Cristian and Kim, Yong Jin and Lombardo, Antonio and Milana, Silvia and Nair, Rahul Raveendran and Novoselov, Konstantin S and Ferrari, Andrea C and Cerullo, Giulio and others},
  journal={Nature communications},
  volume={4},
  number={1},
  pages={1987},
  year={2013},
  publisher={Nature Publishing Group UK London}
}

@article{el2016epitaxial,
  title={Epitaxial graphene quantum dots for high-performance terahertz bolometers},
  author={El Fatimy, Abdel and Myers-Ward, Rachael L and Boyd, Anthony K and Daniels, Kevin M and Gaskill, D Kurt and Barbara, Paola},
  journal={Nature nanotechnology},
  volume={11},
  number={4},
  pages={335--338},
  year={2016},
  publisher={Nature Publishing Group UK London}
}

@article{riccardi2020ultrasensitive,
  title={Ultrasensitive photoresponse of graphene quantum dots in the coulomb blockade regime to THz radiation},
  author={Riccardi, Elisa and Massabeau, Sylvain and Valmorra, Federico and Messelot, Simon and Rosticher, Michael and Tignon, J{\'e}r{\^o}me and Watanabe, Kenji and Taniguchi, Takashi and Delbecq, Matthieu and Dhillon, Sukhdeep and others},
  journal={Nano Letters},
  volume={20},
  number={7},
  pages={5408--5414},
  year={2020},
  publisher={ACS Publications}
}

@article{walsh2021josephson,
  title={Josephson junction infrared single-photon detector},
  author={Walsh, Evan D and Jung, Woochan and Lee, Gil-Ho and Efetov, Dmitri K and Wu, Bae-Ian and Huang, K-F and Ohki, Thomas A and Taniguchi, Takashi and Watanabe, Kenji and Kim, Philip and others},
  journal={Science},
  volume={372},
  number={6540},
  pages={409--412},
  year={2021},
  publisher={American Association for the Advancement of Science}
}

@article{massicotte2021hot,
  title={Hot carriers in graphene--fundamentals and applications},
  author={Massicotte, Mathieu and Soavi, Giancarlo and Principi, Alessandro and Tielrooij, Klaas-Jan},
  journal={Nanoscale},
  volume={13},
  number={18},
  pages={8376--8411},
  year={2021},
  publisher={Royal Society of Chemistry}
}

@article{benford2004mission,
  title={Mission concept for the single aperture far-infrared (SAFIR) observatory},
  author={Benford, Dominic J and Amato, Michael J and Mather, John C and Moseley Jr, S Harvey and Leisawitz, David T},
  journal={Astrophysics and Space Science},
  volume={294},
  number={3},
  pages={177--212},
  year={2004},
  publisher={Springer}
}

@article{graham2015experimental,
  title={Experimental searches for the axion and axion-like particles},
  author={Graham, Peter W and Irastorza, Igor G and Lamoreaux, Steven K and Lindner, Axel and van Bibber, Karl A},
  journal={Annual Review of Nuclear and Particle Science},
  volume={65},
  number={1},
  pages={485--514},
  year={2015},
  publisher={Annual Reviews}
}

@article{fong2012ultrasensitive,
  title={Ultrasensitive and Wide-Bandwidth Thermal Measurements of Graphene<? format?> at Low Temperatures},
  author={Fong, Kin Chung and Schwab, KC},
  journal={Physical Review X},
  volume={2},
  number={3},
  pages={031006},
  year={2012},
  publisher={APS}
}

@article{graham2013photocurrent,
  title={Photocurrent measurements of supercollision cooling in graphene},
  author={Graham, Matt W and Shi, Su-Fei and Ralph, Daniel C and Park, Jiwoong and McEuen, Paul L},
  journal={Nature Physics},
  volume={9},
  number={2},
  pages={103--108},
  year={2013},
  publisher={Nature Publishing Group UK London}
}

@article{block2021observation,
  title={Observation of giant and tunable thermal diffusivity of a Dirac fluid at room temperature},
  author={Block, Alexander and Principi, Alessandro and Hesp, Niels CH and Cummings, Aron W and Liebel, Matz and Watanabe, Kenji and Taniguchi, Takashi and Roche, Stephan and Koppens, Frank HL and van Hulst, Niek F and others},
  journal={Nature Nanotechnology},
  volume={16},
  number={11},
  pages={1195--1200},
  year={2021},
  publisher={Nature Publishing Group UK London}
}

@article{govia2015high,
  title={High-fidelity qubit measurement with a microwave photon counter},
  author={Govia, Luke CG and Pritchett, Emily J and Xu, Canran and Plourde, BLT and Vavilov, Maxim G and Wilhelm, Frank K and McDermott, R},
  journal={arXiv preprint arXiv:1502.01564},
  year={2015}
}

@article{lee2020graphene,
  title={Graphene-based Josephson junction microwave bolometer},
  author={Lee, Gil-Ho and Efetov, Dmitri K and Jung, Woochan and Ranzani, Leonardo and Walsh, Evan D and Ohki, Thomas A and Taniguchi, Takashi and Watanabe, Kenji and Kim, Philip and Englund, Dirk and others},
  journal={Nature},
  volume={586},
  number={7827},
  pages={42--46},
  year={2020},
  publisher={Nature Publishing Group UK London}
}

@article{efetov2018fast,
  title={Fast thermal relaxation in cavity-coupled graphene bolometers with a Johnson noise read-out},
  author={Efetov, Dmitri K and Shiue, Ren-Jye and Gao, Yuanda and Skinner, Brian and Walsh, Evan D and Choi, Hyeongrak and Zheng, Jiabao and Tan, Cheng and Grosso, Gabriele and Peng, Cheng and others},
  journal={Nature nanotechnology},
  volume={13},
  number={9},
  pages={797--801},
  year={2018},
  publisher={Nature Publishing Group UK London}
}

@article{Walsh2021,
  author  = {Walsh, Evan D. and Jung, Woochan and Lee, Gil-Ho and Efetov, Dmitri K. and
             Wu, Bae-Ian and Huang, K.-F. and Ohki, Thomas A. and Taniguchi, Takashi and
             Watanabe, Kenji and Kim, Philip and Englund, Dirk and Fong, Kin Chung},
  title   = {{J}osephson junction infrared single-photon detector},
  journal = {Science},
  volume  = {372},
  number  = {6540},
  pages   = {409--412},
  year    = {2021},
  doi     = {10.1126/science.abf5539}
}

@article{Walsh2017,
  author  = {Walsh, Evan D. and Efetov, Dmitri K. and Lee, Gil-Ho and Heuck, Mikkel and
             Crossno, Jesse and Ohki, Thomas A. and Kim, Philip and Englund, Dirk and
             Fong, Kin Chung},
  title   = {Graphene-based {J}osephson-junction single-photon detector},
  journal = {Physical Review Applied},
  volume  = {8},
  pages   = {024022},
  year    = {2017},
  doi     = {10.1103/PhysRevApplied.8.024022}
}

\end{document}